\documentclass[twocolumn,pra,aps,superscriptaddress,showpacs]{revtex4-1}

 \usepackage{mathptmx}
\usepackage{dcolumn}
\usepackage{amsmath,amssymb,amsfonts}
\usepackage{dsfont}
\usepackage{bm}
\usepackage{color}
\usepackage{latexsym}
\usepackage{epstopdf}
\usepackage[english]{babel}
\usepackage{enumerate}
\usepackage[utf8]{inputenc}

\usepackage[pdftex]{graphicx}

\usepackage{natbib}
\usepackage{epstopdf}
\usepackage{appendix}

\definecolor{mygrey}{gray}{0.35}
\definecolor{myblue}{rgb}{0.2,0.2,0.8}
\definecolor{myzard}{cmyk}{0,0,0.05,0}
\definecolor{mywhite}{rgb}{1,1,1}
\definecolor{mywhite}{rgb}{1,1,1}
\definecolor{myred}{rgb}{1,0.,0.3}

\usepackage[colorlinks=true,citecolor=myblue,linkcolor=myred]{hyperref}

\newcommand{\ket}[1]{\left| #1\right\rangle}

\usepackage[normalem]{ulem}
\usepackage{soul}

\begin{document}

\title{Chiral quantum optics in photonic sawtooth lattices}

\author{Eduardo S\'anchez-Burillo}
\affiliation{Max-Planck-Institut f\"ur Quantenoptik, Hans-Kopfermann-Str. 1, 85748 Garching, Germany}

\author{Chao Wan}
\affiliation{Max-Planck-Institut f\"ur Quantenoptik, Hans-Kopfermann-Str. 1, 85748 Garching, Germany}
\affiliation{Fakult\"at Physik at Ludwig-Maximilians-Universit\"at, Schellingstra{\ss}e 4, 80799 M\"unchen, Germany}

\author{David Zueco}
\affiliation{Instituto de Ciencia de Materiales de Arag\'on and Departamento de F\'isica de la Materia Condensada, CSIC-Universidad de Zaragoza, Calle Pedro Cerbuna 12, 50009 Zaragoza, Spain}
\affiliation{Fundaci\'on ARAID, Paseo Mar\'ia Agust\'in 36, 50004 Zaragoza, Spain}

\author{Alejandro Gonz\'alez-Tudela}
\affiliation{Instituto de F\'isica Fundamental IFF-CSIC, Calle Serrano 113b, Madrid 28006, Spain}

\date{\today}

\begin{abstract}
Chiral quantum optics has become a burgeoning field due to its potential applications in quantum networks or quantum simulation of many-body physics. Current implementations are based on the interplay between local polarization and propagation direction of light in nanophotonic structures. In this manuscript, we propose an alternative platform based on coupling quantum emitters to a photonic \emph{sawtooth} lattice, a one-dimensional model with an effective flux per plaquette introduced by complex tunnelings. We study the dynamics emerging from such structured photonic bath and find the conditions to obtain quasi-perfect directional emission when the emitters are resonant with the band. In addition, we find that the photons in this bath can also mediate complex emitter-emitter interactions tunable in range and phase when the emitters transition frequencies lie within a band-gap. Since these effects do not rely on polarization they can be observed in platforms beyond nanophotonics such as matter-waves or circuit QED ones, of which we discuss a possible implementation.
\end{abstract}

\maketitle

\emph{Introduction.- } Designing robust non-reciprocal optical devices at the classical and quantum level has remained a challenge for many years with fundamental and practical implications~(see, e.g., Refs.~\cite{jalas13a,sayrin15a,sounas17a,caloz18a}, and references therein). On the classical level, the search was mainly focused on systems breaking Lorentz-reciprocity such as magneto-optical materials. On the quantum side, nanophotonic systems have recently emerged as a powerful candidate due to the link between the local polarization and propagation direction that appears thanks to the subwavelength light confinement~\cite{bliokh15a,vanmechelen16a}. Using this connection and the intrinsic polarization of classical and quantum emitters (nanoparticles, atoms, or quantum dots) many experiments have shown chiral light-matter coupling in waveguides~\cite{mitsch14a,Petersen2014,Scheucher2016,sollner15a,coles16a}, and harnessed it to achieve, e.g., optical isolation at the single-photon level~\cite{sayrin15a}. Apart from these experiments, chiral quantum optical systems~\cite{Lodahl2017b} have also been proposed to build spin-photon interfaces~\cite{mahmoodian16a} and directional amplifiers~\cite{metelmann15a,malz18a} in quantum networks, to engineer novel many-body spin or photonic states~\cite{ramos14a,ramos16a,vermersch16a,pichler15a,guimond16a}, to emit non-classical light~\cite{mahmoodian18a,downing19a}, or to induce exotic self-organization patterns~\cite{eldredge16a}, among other phenomena. 

\begin{figure}[tb!]
 \includegraphics[width=0.4\textwidth]{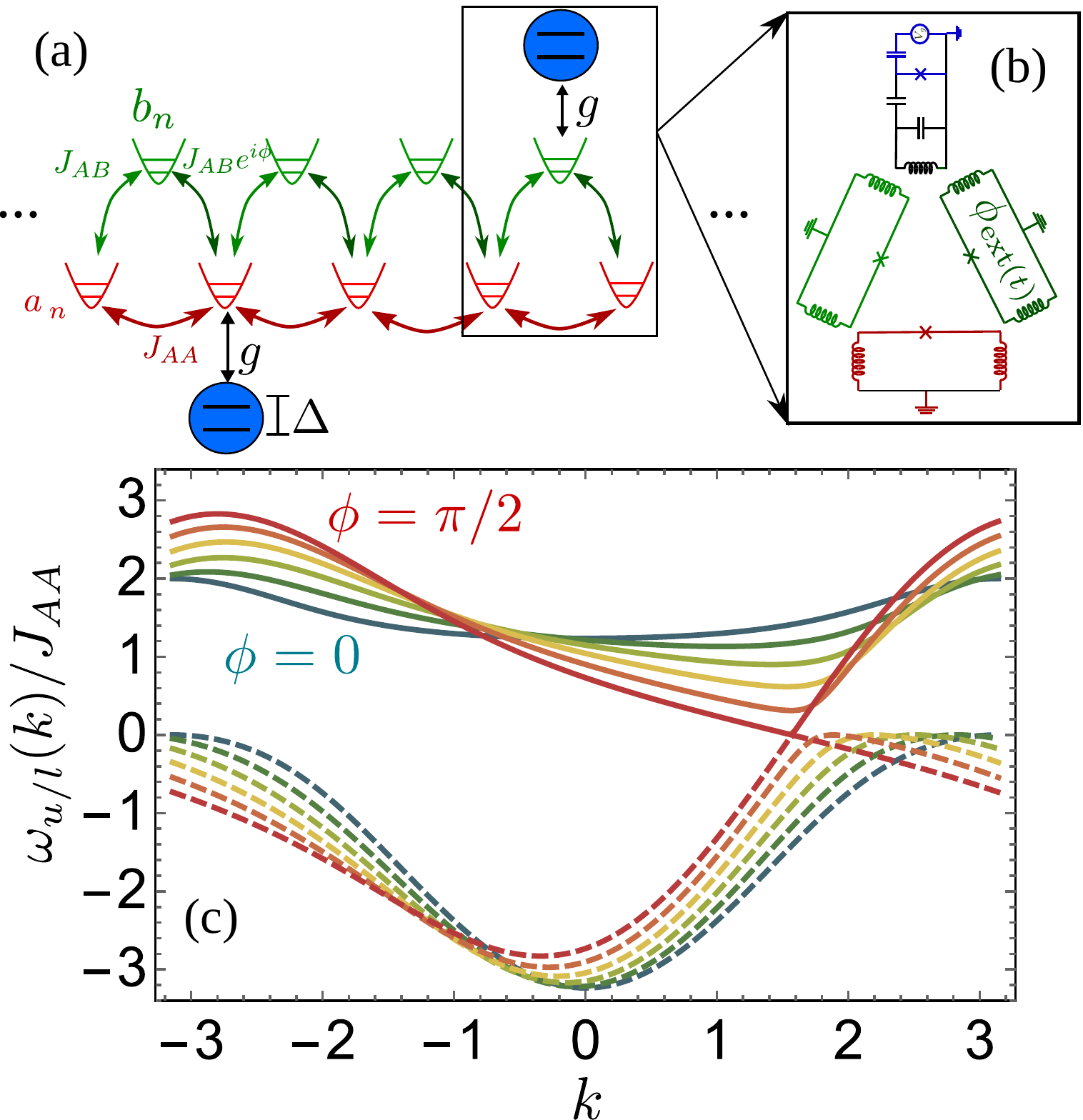}
    \caption{(Color online). (a) Scheme of two QEs (in blue), with transition frequency $\Delta$, coupled locally with strength $g$ to a photonic sawtooth lattice. The red/green lattice sites denote the A/B sublattices which are coupled in a loop with rates $J_{AA}, J_{AB}$ and $J_{AB} e^{i\phi}$ as depicted in the figure. (b) Sketch of the superconducting circuit to simulate a sawtooth lattice: A superconducting qubit (in blue) coupled capacitively to an LC resonator (in black), which mediates the interaction with the resonators of the lattice, also LC (in green and red). The LC resonators are coupled themselves via Josephson junctions, whose external flux can be controlled externally in a time-dependent fashion. (c)  Photonic bands $\omega_{u/l}(k)$ (solid/dashed) for $J_{AA}=J_{AB}$, for six equally spaced $\phi$ from $\phi=0$ (dark green) to $\phi$ and $\phi=\pi/2$ (red).}
    \label{fig:sketch}
\end{figure}

All these exciting prospects have triggered the interest to implement these chiral couplings also beyond the optical regime. There are theoretical proposals to do so using complex non-local light-matter interactions~\cite{ramos14a,ramos16a,vermersch16a} or with moving atoms~\cite{calajo17a}, but their experimental realization remains so far elusive. In this manuscript, we explore an alternative in which quantum emitters (QEs) couple locally to a photonic bath which inherently breaks time-reserval symmetry. In particular, we use a minimal one-dimensional model named as photonic sawtooth lattice (see Fig.~\ref{fig:sketch}(a)), also labelled as $\Delta$ chain \cite{Nakamura1996} or triangle lattice \cite{Hyrkas2013,Flach2014} in the literature. This is a one-dimensional model with closed loops, which allows for a complex coupling ($\phi\neq 0$ in Fig.~\ref{fig:sketch}) between the degrees of freedom in the photonic lattice vertices defining an effective magnetic flux per loop~\footnote{If there are no loops, complex couplings have nonphysical implications.}. We first propose an implementation to simulate this model based on circuit QED technologies~\cite{Astafiev2010,Hoi2011,Hoi2013a,VanLoo2013,Hoi2013b,liu17a,mirhosseini18a,Roushan2016,Gu2017}.
We then predict that one can indeed obtain quasi-perfect directional emission~\cite{Lodahl2017b} when the QEs are resonant with the bands appearing in these systems. In our proposal, rather than selecting a given momentum using the destructive interference induced by non local couplings~\cite{ramos14a,ramos16a,vermersch16a}, it is the asymmetric nature of the ``sawtooth'' band structure the one responsible of the chirality of the emission. In addition to that, the properties of the bath lead to other phenomena such as the emergence of a sublattice-dependent directional emission or QE dipole interactions tunable in range an phase when the QEs frequencies lie in a band-gap.

\emph{Model and implementation.- }The Hamiltonian of the full system (photons and QEs) is ($\hbar=1$):
\begin{equation}\label{eq:H}
    H=H_\text{ST} + \Delta  \sum_{j=1}^M \sigma_j^\dagger \sigma_j + H_\text{int},
\end{equation}
where $\Delta$ is the transition frequency of each QE described as a two-level system with ladder operators $\sigma_j^{(\dagger)}$ for the $j$-th QE, $M$ is the number of QEs, $H_\text{int}$ is the interaction Hamiltonian, and $H_\text{ST}$ is the Hamiltonian of the sawtooth lattice. The latter reads:
\begin{align}\label{eq:HST}
H_\text{ST} = & \omega_{B} \sum_{n=1}^N (a_n^\dagger a_n + b_n^\dagger b_n) -  J_{AA}\sum_{n=1}^N (a_n^\dagger a_{n+1} + \text{H.c.}) \nonumber\\
&- J_{AB} \sum_{n=1}^N (a_n^\dagger b_n + e^{-i\phi} a_{n+1}^\dagger b_n + \text{H.c.}),
\end{align}
where $a_n$ and $b_n$ are the annihilation operators of the $n$-th $a$ and $b$ modes (notice that the photonic bath is bipartite and must be described by two sublattices $A$ and $B$, see Fig.~\ref{fig:sketch}(a)), $N$ is the number of modes in each sublattice, $\omega_B$ is the bare energy of each resonator that we take as the energy reference, i.e., $\omega_B\equiv 0$ , $J_{AA}$ is the nearest-neighbour coupling between the $A$ lattice sites, $J_{AB}$ stands for the hopping strength between the $a$ and $b$ modes, and $\phi$ is the phase difference in each loop (see Fig.~\ref{fig:sketch}(a)).

As the physics of Eq.~\eqref{eq:HST} does not rely on polarization, it can be obtained in platforms beyond optical ones~\cite{mitsch14a,Petersen2014,Scheucher2016,sollner15a,coles16a} such as cold atoms in state-dependent lattices~\cite{devega08a,navarretebenlloch11a,krinner18a}, where such complex loops have already been engineered~\cite{aidelsburger13a}, or circuit QED platforms~\cite{Astafiev2010,Hoi2011,Hoi2013a,VanLoo2013,Hoi2013b}. In Fig.~\ref{fig:sketch}(b) we show a way on how to engineer such looped geometries and complex phases using superconducting qubits as implemented for a single loop in Ref.~\cite{Roushan2016}. For the qubit-resonator coupling, we assume that the qubit is capacitively coupled to an LC resonator as shown in Fig.~\ref{fig:sketch}(b), whereas the resonators are inductively coupled via a SQUID-type loop based on Josephson junctions~\cite{Peropadre2013a, Chen2014}. These Josephson junctions can be controlled via an external flux $\phi_{\rm ext}(t)$ entering the loop leading to the time-dependent coupling Hamiltonian between two resonators:
\begin{equation}
    H_\text{aux} = \sum_{i=1,2} \omega_i a_i^\dagger a_i
    + J(t) (a_1^\dagger + a_1 ) (a_2^\dagger + a_2)
\end{equation}

Then, setting $\omega_1= \omega$ and $\omega_2= \omega + \delta$, taking $J(t) = J \cos (\delta t + \phi) $, and assuming $J, \delta\ll\omega$, we arrive to the effective Hamiltonian we want to simulate, cf.~Eq.~\eqref{eq:HST}. For the experimentally implemented loop of Ref.~\cite{Roushan2016}, couplings of the order of $4.1$ MHz (i.e. $J / \omega \sim 10^{-3}$) have been measured. Thus, concatenating several of these cells and placing selectively the qubits, as done in Refs.~\cite{liu17a,mirhosseini18a} for simple coupled cavity arrays, one will be able to explore the phenomena predicted in this manuscript.

For the interaction term $H_\text{int}$ (last term of Eq.~\eqref{eq:H}), we consider point-like and dipole-field coupling under the rotating-wave approximation, which is valid when the coupling strength is small with respect to the other energy scales of the system \cite{Cohen-Tannoudji1992}: 
\begin{equation}\label{eq:Hint}
H_\text{int} = g\left(\sum_{j=1}^{M_a}\sigma_j^+ a_{x_j} + \sum_{j=M_a+1}^{M}\sigma_j^+ b_{x_j}\right) + \text{H.c.}
\end{equation}
Here $g$ is the coupling constant, $M_a$ is the number of qubits coupled to $A$ (so $M_b\equiv M-M_a$ are coupled to $B$), and $x_j$ is the position of the $j$-th qubit.

Since we are interested in predictions in the thermodynamic limit, that is, when $N\rightarrow \infty$, we can take periodic boundary conditions for the bath and introduce the following plane wave modes $\hat{a}_k/\hat{b}_k\equiv 1/\sqrt{N}\sum_{n=1}^N e^{-i k n}a_n/b_n$, in terms of which $H_\text{ST}$ (Eq. \eqref{eq:HST}) reads:
\begin{equation}\label{eq:HST_fourier}
    H_\text{ST} = \sum_k \left(\hat{a}_k^\dagger\; \hat{b}_k^\dagger\right) h_\text{ST}(k) \left(\begin{array}{c}
         \hat{a}_k  \\
         \hat{b}_k 
    \end{array}\right),
\end{equation}
with $h_\text{ST}(k)$
\begin{equation}
      h_\text{ST}(k) =  \left(
    \begin{array}{cc}
        -2J_{AA}\cos k & f(k,\phi)  \\
        f^*(k,\phi) & 0
    \end{array}
    \right),
\end{equation}
where $f(k,\phi)= -J_{AB}(1+e^{-i(k+\phi)})$. Notice that we introduce the $\hat{\cdot}$ notation to distinguish the operators in real/momentum space. We can diagonalize $h_\text{ST}(k)$ such that $H_\text{ST} = \sum_k (\omega_u(k) \, \hat{u}_k^\dagger \hat{u}_k + \omega_l(k) \, \hat{l}_k^\dagger \hat{l}_k)$, where $\hat{u}_k$ and $\hat{l}_k$ are related to $\hat{a}_k$ and $\hat{b}_k$ by means of a unitary transformation:
\begin{equation}\label{eq:modes}
\left(\begin{array}{c}
    \hat{u}_k \\
     \hat{l}_k
\end{array}\right) = 
\left(\begin{array}{cc}
    \cos(\theta_k)e^{-i\varphi_k} & -\sin(\theta_k) \\
     \sin(\theta_k)e^{-i\varphi_k} & \cos(\theta_k)
\end{array}\right)
\left(
\begin{array}{c}
    \hat{a}_k  \\
    \hat{b}_k
\end{array}\right),
\end{equation}
where the particular form of $\theta_k$ and $\varphi_k$ is shown in the Sup. Material \cite{SupMat}.
The bands of the model, $\omega_{u/l}(k)$, read
\begin{align}\label{eq:bands}
    \omega_{u/l}(k) = -J_{AA}\cos k \pm \sqrt{J_{AA}^2\cos^2 k+4 J_{AB}^2\cos^2\left((k+\phi)/2\right)}.
\end{align}

A celebrated feature of the sawtooth lattice is the appearance of flat bands, setting $J_{AB}/J_{AA}=\sqrt{2}$ and $\phi=0$ \cite{Leykam2018}. Here, we are however interested in the  implications of a nontrivial phase, $\phi\neq 0$, which leads nonsymmetric band structure in $k$-space (see Fig.~\ref{fig:sketch}(c)), due to the explicit breaking of time reversal ($H_\text{ST}\neq H_\text{ST}^*$, which implies $h_\text{ST}(k)\neq h_\text{ST}(-k)$).
 This cannot happen in a standard 1D photonic system (without loops) since one can get rid of the phase $\phi$ by means of local transformations of the bosonic operators. Apart from the $k$-asymmetry, there appears an extra band-gap between both bands for most values of $\phi$, except for $\phi=\pm\pi/2$ (see again Fig.~\ref{fig:sketch}(c)) where the two bands touch at single $k$-points $k=\pm\pi/2$. These type of singular bandgaps lead to exotic QE dynamics and interactions in higher dimensions~\cite{Tudela2018a,perczel18a,gonzaleztudela18d}. This will not be the case for the bath considered in this manuscript since the coupling strength to the upper/lower band $\omega_{l/u}(k)$, defined by the functions $\theta_k,\varphi_k$, turn this point into a trivial band-crossing, as we explain in Sup.~Material~\cite{SupMat}.

\emph{QE resonant with the band: Directional emission.- } We first study the spontaneous decay of a single qubit ($M=1$ in \eqref{eq:H} and \eqref{eq:Hint}) when the QE transition frequency lies within the bands, i.e., $\Delta\in \omega_{u/l}(k)$. The state at time $t$ reads $|\Psi_D(t)\rangle = e^{-iHt} \sigma^\dagger \ket{\mathrm{vac}}$, where $\ket{\mathrm{vac}}$ is the global vacuum state and $D$ stands for the sublattice the QE is coupled to: $D=A,B$. As the number of excitations $\mathcal{N}\equiv\sum_{n=1}^N(a_n^\dagger a_n + b_n^\dagger b_n) + \sigma^\dagger \sigma$ is a conserved quantity under the rotating-wave approximation \eqref{eq:Hint}, the state $|\Psi_D(t)\rangle$ can be spanned in the single-excitation subspace as
\begin{equation}
    |\Psi_D(t)\rangle = \left(c_e^D(t)\sigma^+ + \sum_k \left(c_u^D(k,t) \hat{u}_k^\dagger + c_l^D(k,t) \hat{l}_k^\dagger \right)\right)|0\rangle.
\end{equation}

\begin{figure}
	\centering
	\includegraphics[scale=0.28]{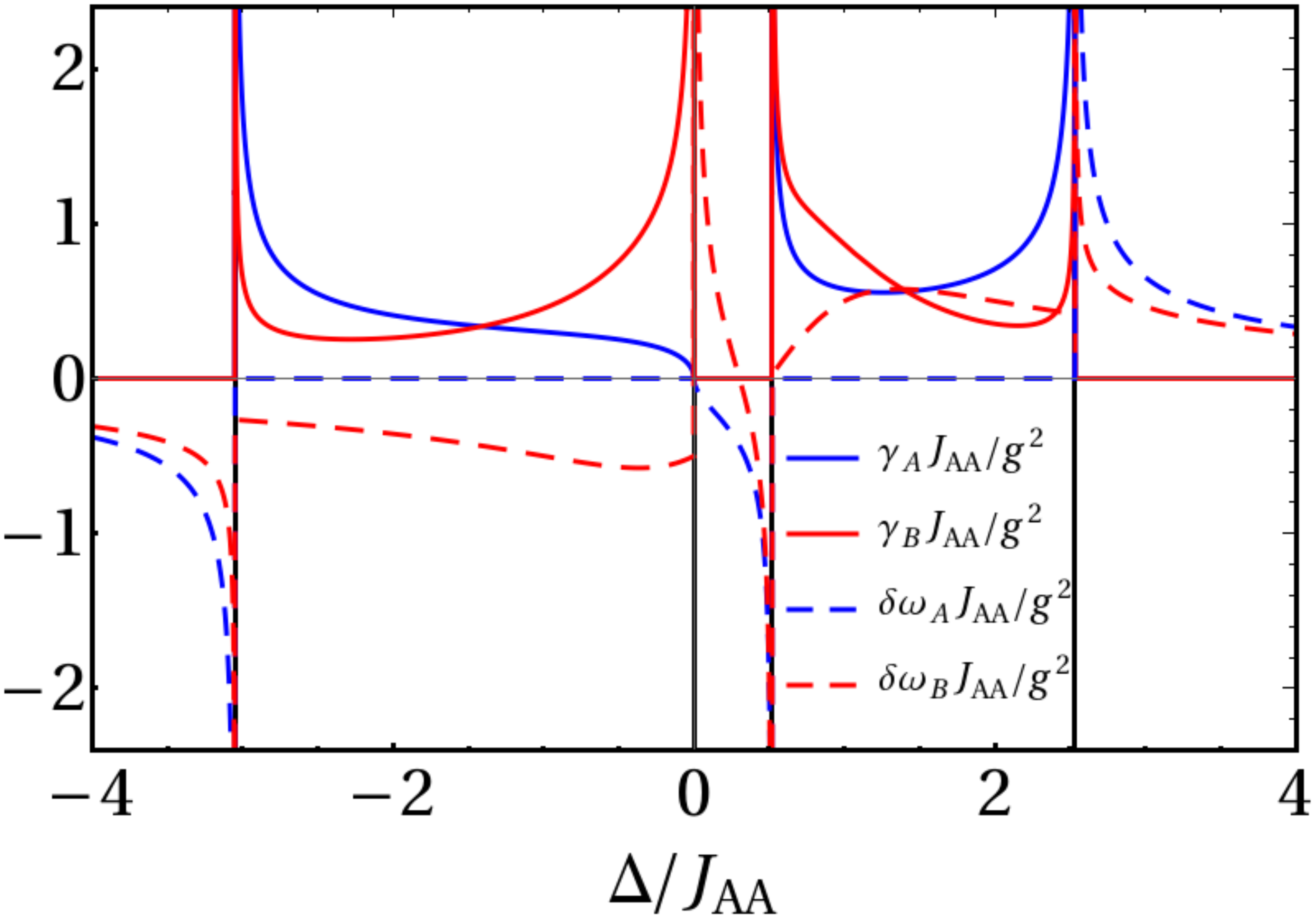}
	\caption{(Color online). Decay rate $\gamma_D = -2\text{Im}\Sigma_e^D$ (solid lines) and energy shift $\delta\omega_D(\Delta)= \text{Re}\Sigma_e^D(\Delta+i0^+)$ (dashed lines) as a function of the excitation energy of the qubit for the $A$ (blue) and $B$ (red) sublattices. The parameters are $J_{AB}=J_{AA}$ and $\phi=\pi/3$. The vertical black lines stand for the band limits.}
	\label{fig:self_energy_1qb}
\end{figure}

This allows us to calculate exactly the dynamics either numerically or semi-analytically using the resolvent operator approach~\cite{Cohen-Tannoudji1992}. In the latter, the QE probablity amplitude, $c_e^D(t)$, is obtained as the inverse Laplace transform of the QE Green Function operator~$c_e^D(t) = {\mathcal L}^{-1} [1 / (s + \Sigma_e^D(z))]$
where the self-energy reads~\cite{SupMat}:
\begin{equation}\label{eq:sigmaD}
\Sigma_e^D(z)=\sum_k \sum_{\alpha=u,l}\frac{|\langle 0 |\alpha_k H_\text{int}\sigma^+|0\rangle|^2}{z-\omega_\alpha(k)}\,.
\end{equation}

Within the Markovian approximation the dependence on $z$ of the self-energy is neglected and replaced by $z=\Delta+i 0^+$ when doing the inverse Laplace, yielding $c_e^D(t)\simeq e^{-i\Sigma_e^D(\Delta+i0^+)t}$, where $\Sigma_e^D(\Delta + i0^+)=\delta\omega_D-i\gamma_D/2$. Therefore, $\delta \omega_D$ and $\gamma_D$
correspond to the renormalization of the excited state frequency and linewidth, respectively. In Fig.~\ref{fig:self_energy_1qb} we plot these functions, $\delta\omega_D$ and $\gamma_D$, for $\phi=\pi/3$, marking the four band limits with vertical lines.  The main result here is that both the decay rate and Lamb-shift depends on the sublattice the QE is coupled, which is a consequence of sublattice symmetry breaking of the bath for any $J_{AB}\neq 0$. When $\phi\neq \pi/2$, the latter also leads to the cancellation of the standard 1D band-edge divergence of $\gamma_A$ when $\Delta$ matches the the upper [lower] band-edge of $\omega_l(k)$ $[\omega_u(k)]$ for $\phi\in [0,\pi/2)$ [$(\pi/2,\pi]$]. Similar behaviour was also found in two-dimensional photonic crystals without sublattice symmetry~\cite{gonzalez18f}.

\begin{figure}
    \centering
    \includegraphics[scale=0.26]{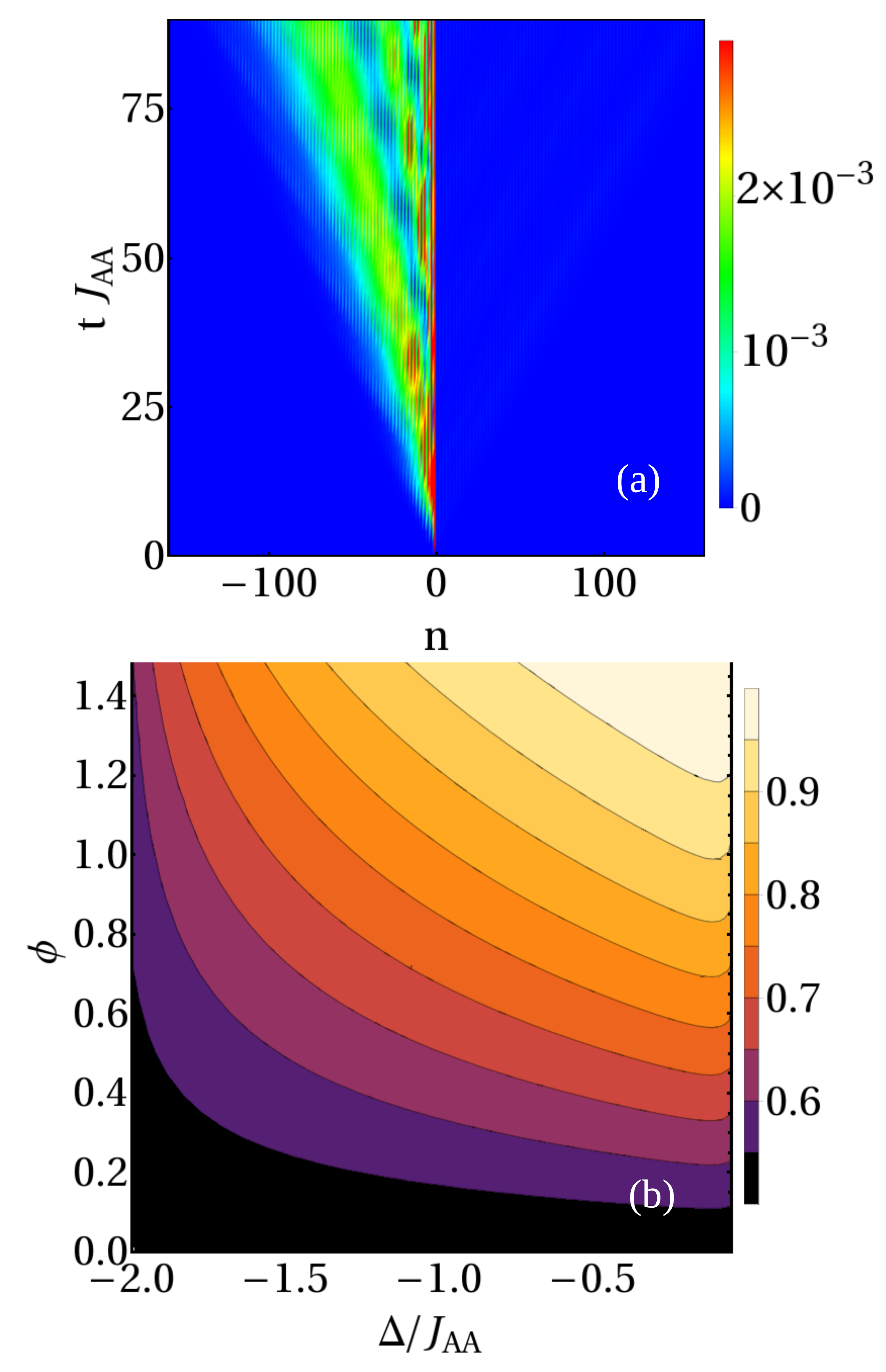}
    \caption{(Color online). (a) Emitted photon in real space as a function of time, $\langle a_n^\dagger a_n\rangle(t)$ and $\langle b_n^\dagger b_n\rangle(t)$, when the qubit is coupled to $B$ for $J_{AB}=0.2 J_{AA}$, $\phi=1.5$, $\Delta=-0.5 J_{AA}$, $g=0.1 J_{AA}$, and bath size $N=200$. The qubit is placed at $n=0$. (b) Directionality ratio $R^B_L$ for $J_{AB}=0.2 J_{AA}$ as a function of $\Delta$ and $\phi$.}
    \label{fig:directionality}
\end{figure}

When calculating the exact QE dynamics (not shown), we find band-edge related dynamics such as fractional and power-law decays in the long-time limit when the atomic frequencies are close to the band edges, which are very similar to the ones appearing in other band-gap photonic materials~\cite{Khalfin1958,Bykov1975,Fonda1978,Hack1982,Onley1992,John1994,Gaveau1995,Garmon2013,Redchenko2014,Lombardo2014,Sanchez-Burillo2016a,gonzaleztudela18c,gonzalez18f}. Thus, for this manuscript, we focus on the bath emission dynamics which indeed displays very distinctive features from other photonic baths.  As a first evidence of that, we plot in Fig.~\ref{fig:directionality}(a) a snapshot of the photon population in real space for a situation where the emission is highly directional, which corresponds to a QE coupled to the $B$ sublattice, with parameters $J_{AB}=0.2 J_{AA}$, $\phi=1.5$, and $\Delta=-0.5 J_{AA}$. We want to note that even though the bath breaks $\pm k$ symmetry for any $\phi\neq 0$, we numerically find that the degree of directional emission depends strongly on other parameters, especially $\Delta$. Thus, having a bath with $\phi\neq 0$ is therefore a necessary but not sufficient condition for efficient directional emission.

Let us further understand the origin and possibilities of the directional emission in this system by considering that the QE is resonant with $\omega_l(k)$, and taking the limit when $g$ is small enough such that we are in the Markov regime. In this limit, the QE dynamics is dominated by the resonant $k$-modes defined by $\omega_l(k_{R/L})=\Delta$, where $k_{R/L}$ correspond to the momenta of the right- and left-moving photons, respectively (see Fig.~\ref{fig:sketch}(b)). Furthermore, within each direction the excitations split between the ones propagating in the $A$ or $B$ sublattices. Thus, the decay rate $\gamma_D$ introduced as the imaginary part of the self-energy (Eq.~\eqref{eq:sigmaD}) can be separated into four different contributions:
\begin{align}
    &\gamma_D=\Gamma^D_{a}(k_R)+\Gamma^D_{a}(k_L)+\Gamma^D_{b}(k_R)+\Gamma^D_{b}(k_L)\,,
     \label{eq:gammaB}
    \end{align}
where $\Gamma^D_{\alpha}(q)$ denotes the decay rate into the $\alpha$-sublattice at momentum $q$ for a QE coupled to the $D$ sublattice, reading:
\begin{align}
    \Gamma^{A(B)}_{a(b)}(q)&=\frac{|\sin(\theta_q)|^4\left(|\cos(\theta_q)|^4\right)}{|v_l(q)|}\label{eq:lambdaa}\\
    \Gamma^{A}_{b}(q)&=\frac{|\sin(\theta_q)|^2|\cos(\theta_q)|^2}{|v_l(q)|}=\Gamma^{B}_{a}(q) ~\label{eq:lambdab}
\end{align}
where $v_l(k)$ is the group velocity of the modes in the lower band, $v_l(k)=\partial_k \omega_l(k)$. With these functions we can define a global directionality ratio:
\begin{align}\label{eq:directionality_ratio}
R_{L/R}^D=\frac{\sum_{\alpha}\Gamma^D_\alpha(k_{R/L})}{\sum_{\alpha}\left(\Gamma^D_{\alpha}(k_{R})+\Gamma^D_{\alpha}(k_{L})\right)}
\end{align}
with $\alpha=a,b$, that tell us the ratio of light emitted in the left/right side in both sublattices, and a local one which distinguishes between sublattices $R^{D}_{R/L,a/b}$ with the same expressions but without the summation in $\alpha$. 

In Fig.~\ref{fig:directionality}(b), we plot $R^B_L$ as a function of $\Delta$ and $\phi>0$ for a tunneling $J_{AB}=0.2 J_{AA}$. Note that the role of the $L/R$ is reversed by switching the sign of $\phi$. There, we observe that we can find non-reciprocal emission, that is, $R_\alpha^D>1/2$ for any $\phi\neq 0$. However, in order to find $R^B_L\approx 1$ one has to go to the limit where $J_{AB}/J_{AA}\ll 1$, $\phi\rightarrow \pi/2$, and $\Delta\to 0^-$ where the slope around $k=\pm\pi/2$ is very different yielding a highly asymmetric density of states for the left/right moving modes. The left/right character of the emission can be switched with the sign of $\phi$. When the QE couples to the $A$ sublattice instead, the global $R^A_{L/R}\approx 1/2$, however, locally in each sublattice can be made very directional, $R_{L,b}^A,R_{R,a}^A\approx 1$  (see \cite{SupMat}). This is possible because in that case $\theta_{k_{R/L}}$ is such that $\Gamma_a^A(k)$ is drastically much larger for $k_R$, compensating the density of states. To our knowledge, this sublattice-dependent chirality does not appear in other photonic reservoirs considered in the literature.

\emph{QEs outside of the band: Tunable complex interactions.- }We focus now on the regime where $\Delta\notin \omega_{l/u}(k)$, such that the physics is dominated by the bound states (BSs)~\cite{John1984,John1987,Tong2010a,Tong2010b,Longo2010,Longo2011,Yang2013,Lu2013,Sanchez-Burillo2014,Sanchez-Burillo2016a,Calajo2016,Calajo2016b,Shi2016,Shi2018b,Bello2018}. In the single-excitation subspace, the BS wavefunction of a single emitter coupled to the $D$ sublattice reads:
\begin{equation}\label{eq:bs}
    \ket{\Psi_m^D} = \sum_n (c_{m,a}^D(n)a_n^\dagger + c_{m,b}^D(n)b_n^\dagger)\ket{0} + c_{m,e}^D\sigma^\dagger \ket{0}.
\end{equation}
where $m=-1,0,1$ denotes the different BSs that can appear in the upper/middle/lower band-gap, respectively. Their wavefunction, $c_{n,a/b}^D(n)$, and energy can be found from the secular equation: $H \ket{\Psi_m^D}=E_m^D\ket{\Psi_m^D}$ with $E_m^D\notin \omega_{l,u}(k)$ (see SM ~\cite{SupMat}). Using this equation, we can prove that there always exists one bound state $\ket{\Psi_{\mp 1}^D}$ below [above] $\omega_{l[u]}(k)$, because the self-energy always diverges at these band-edges~\cite{shi16a,calajo16a}, such that the interaction with the bath is able to push one state out of the band. However, in the middle band-gap that appears when $\phi\neq \pm \pi/2$, an extra BS $\ket{\Psi_{0}^D}$ emerges if and only if (i) $\Delta>0$, $|\phi|<\pi/2$ or  (ii) $\Delta<0$ and $\phi\in(-\pi,-\pi/2)\cup (\pi/2,\pi)$ for $D=B$ or $A$, respectively~\cite{SupMat}. The underlying reason of this condition is the finite value of the self-energy in one of the band-edges, as shown in Fig.~\ref{fig:self_energy_1qb}(a), which defines a critical detuning for the existence of the BS.

\begin{figure}[tb]
    \centering
    \includegraphics[scale=0.29]{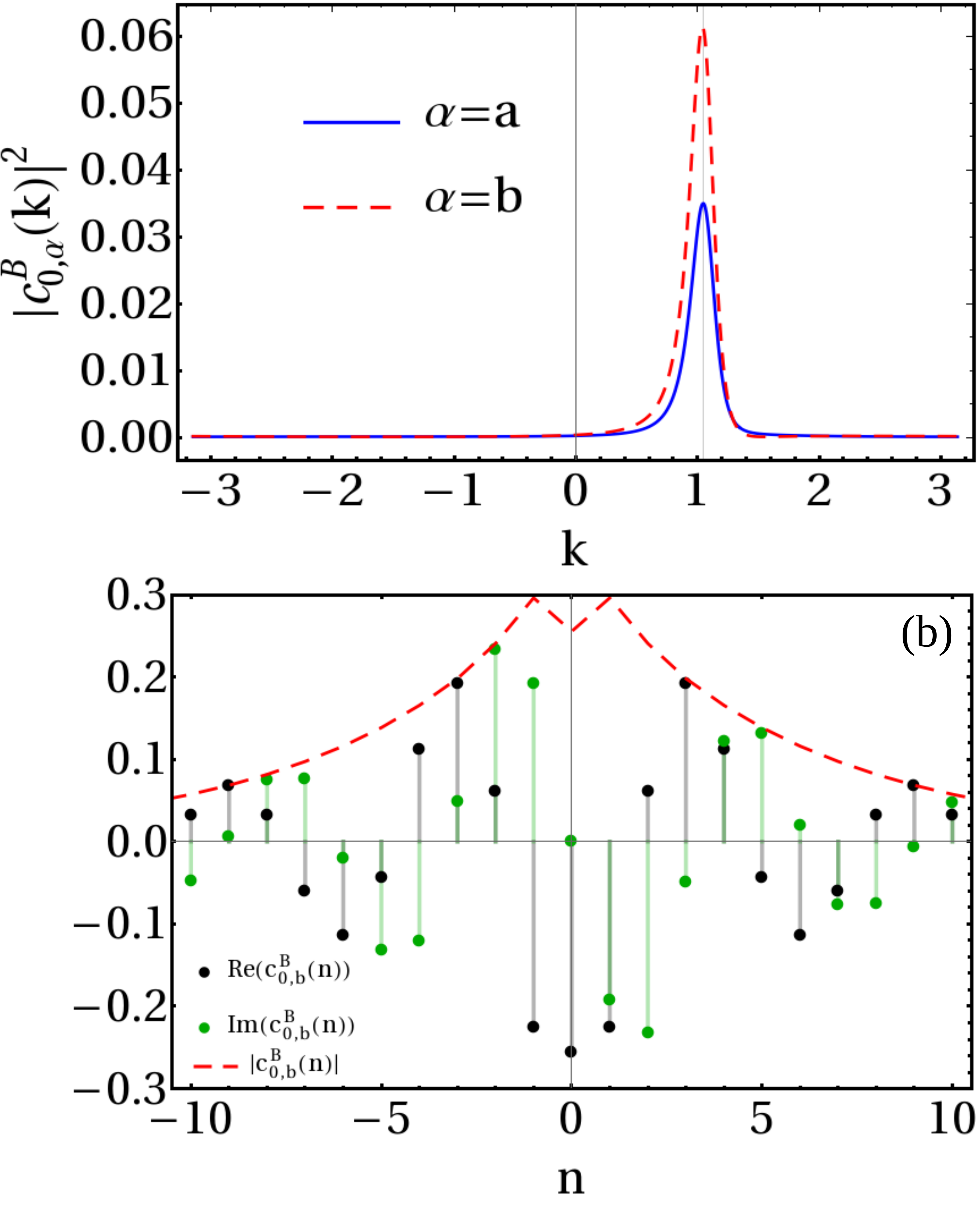}
    \caption{(Color online). Interband bound state $\ket{\Psi_0^B}$ for a qubit coupled to $B$ for $J_{AA}=J_{AB}=1$, $\phi=2.094$, $\Delta=-0.01$, and $g=0.1$. In panel (a) we plot the square modulus of the wavefunctions in momentum space in both sublattices, $|c_{0,a/b}^B(k)|^2$, whereas we show the absolute value (red dashed line) and the real (black dots) and imaginary (green dots) parts of the coefficients in real space for the $b$ modes in panel (b). Notice that the wavefunctions in $k$ space are centered at $k=\pi/3$, so the coefficients in real space have a periodicity of 6.}
    \label{fig:bound_state_mid}
\end{figure}

To illustrate the main features of these BSs, we plot in Fig.~\ref{fig:bound_state_mid} their wavefunction coefficients both in momentum and real space of an instance of a BS; concretely, we choose the interband one $\ket{\Psi_0^D}$. i) Contrarily to what happens in emission, the absolute value of the wavefunction $|c^D_{m,\alpha}(n)|$ is always symmetrically distributed around the QE no matter the band-gap or parameters considered (see Fig.~\ref{fig:bound_state_mid}(b)); ii) as it occurs with other photonic lattices~\cite{John1984,John1987,Tong2010a,Tong2010b,Longo2010,Longo2011,Yang2013,Lu2013,Sanchez-Burillo2014,Sanchez-Burillo2016a,Calajo2016,Calajo2016b,Shi2016,Shi2018b,Bello2018}, the BS are exponentially localized around the emitter with a localization length which can be tuned: the closer the $E^D_{m}$ lies to one of the band-edges, the larger is the localization length; iii) the main distinctive feature is that the BSs acquire a tunable complex phase $c^D_{m,\alpha}(n)\propto e^{i \varphi^D_m n}$. In the small coupling limit, this phase $\varphi^D_m$ matches the position of the band edge closest to $\Delta$. The positions of these band edges change with both the ratio $J_{AB}/J_{AA}$ and $\phi$. This dependence is seen in Fig.~\ref{fig:sketch}(c); e.g., the minimum of the upper band runs from 0 to $\pi/2$, so the phase of the interband bound state can be tuned in this range provided $\Delta$ tends to this band edge. In the example of Fig.~\ref{fig:bound_state_mid}, the momentum of the closest band edge occurs at $k_\mathrm{edge}\simeq\pi/3$, so the coefficients in momentum space $|c_{0,\alpha}^B(k)|^2$ are distributed around $\pi/3$ (panel (a)) and their real and imaginary parts in real space for the $b$ modes, $\text{Re}(c_{0,b}^B(n))$ and $\text{Im}(c_{0,b}^B(n))$ have periodicity $2\pi/(\pi/3)=6$ (panel (b)).

One of the main interests of these BS is that when many emitters couple to the bath, they can mediate interactions between QEs which can be harnessed to simulate spin models with tunable interactions. In the Markovian approximation, an effective Hamiltonian for the qubits can be derived~\cite{douglas15a,gonzaleztudela15c}: 
\begin{align}\label{eq:manybody}
H_\text{qb} = \sum_{i<j}(J^{D_i D_j}_{ij} \sigma_i^\dagger \sigma_j + \text{H.c.})\,,
\end{align}
where $J^{D_i D_j}_{ij} $ inherits the shape of the BS wavefunction with energy $\Delta$, i.e., $J_{ij}^{AA/BB}\propto c_{a/b}^{A/B}(r_{ij})$ and $J^{AB}_{ij}\propto c^A_b(r_{ij})$ (see SM \cite{SupMat}). When $\Delta$ lies in a band-gap, one can then control not only the effective range of the interactions, but, as discussed in the previous paragraph and in Fig.~\ref{fig:bound_state_mid}, also its phase.
In particular, the physical phase of the photonic lattice $\phi$ is inherited by the effective spin-spin interactions (see SM \cite{SupMat}). 
We finally want to note that even richer many-body dynamics will appear in the non-perturbative regime replacing spins by polaritons~\cite{shi18a}.

\emph{Conclusions.- } We have studied the dynamics and interactions of quantum emitters coupled to a minimal one-dimensional model breaking time-reversal symmetry, i.e., the photonic sawtooth lattice. We have found that when the emitters are resonant with the band they decay in an asymmetric fashion into left/right moving modes. Optimizing the parameters we identified regimes of quasi-perfect directionality, or more exotic ones in which the emitter decays in both directions but to a different sublattice in each of the them. Thus, these systems can be an alternative way of exploring chiral quantum optics without the need of using polarization or moving emitters. In addition to that, when the emitter frequency lies in a band-gap we have found the emergence of bound states whose not only their spatial range, but also their complex phase can be tuned through the system parameters. Since these bound states ultimately mediate interactions between emitters when many of them couple to the bath, our proposed setup provides access to the simulation of a large class of spin models with complex interactions. Furthermore, we discussed a particular implementation where to observe such phenomenology based on state-of-the-art superconducting technologies.

\emph{Acknowledgments.- } ESB acknowledges ERC Advanced Grant QUENOCOBA under the EU Horizon 2020 program (grant agreement 742102). AGT and DZ acknowledge support from CSIC Research Platform PTI-001. AGT acknowledges funding from the national project PGC2018-094792-B-I00 from Ministerio de Ciencia e Innovación.


\newpage

\begin{widetext}
	\widetext
	\onecolumngrid
	\begin{center}
		\textbf{\large Supplemental Material: Chiral quantum optics in photonic sawtooth lattices}
	\end{center}
	\vspace{\columnsep}
	\vspace{\columnsep}
	
	\twocolumngrid
\end{widetext}
\setcounter{equation}{0}
\setcounter{figure}{0}
\setcounter{section}{0}
\makeatletter

\renewcommand{\thefigure}{SM\arabic{figure}}
\renewcommand{\thesection}{SM\arabic{section}}  
\renewcommand{\theequation}{SM\arabic{equation}}  

In this Supplementary Material, we give more details of the: i) diagonalization of the bath Hamiltonian in Section~\ref{app:diagonalization}; ii) calculation of the single emitter self-energy in Section~\ref{app:single_qubit_selfenergy}; iii) the structure of the qubit-band couplings in Section~\ref{app:qubit_band}; iv) the absence of a nonanalyticity in the bands for $\phi=\pi/2$; v) more details in the sublattice directional behaviour in Section~\ref{SM:sublat}; vi) the features of the bound states energies and wavefunctions in Section~\ref{app:bs}; vii) the connection between the bound states and the two emitter self-energy in Section~\ref{app:two_qubit_selfenergy}.

\section{Diagonalization of the sawtooth lattice}\label{app:diagonalization}

In this Section, we give some details on the diagonalization of the sawtooth Hamiltonian (Eq. \eqref{eq:HST}). The bosonic operators which diagonalize the model, $u_k/l_k$ (see Eq. \eqref{eq:modes}) are related to $a_k$ and $b_k$ by means of a unitary transformation $P_k$ (Eq. \eqref{eq:modes}). The latter reads
\begin{align}\label{eq:Pk}
    P_k &= 
    \left(
\begin{array}{cc}
    \cos(\theta_k) e^{i\varphi_k}& \sin(\theta_k) e^{i\varphi_k}\nonumber \\
    -\sin(\theta_k) & \cos(\theta_k)
\end{array}
\right)\\
    &=\left(
\begin{array}{cc}
    N_u(k)f^*(k,\phi) & N_l(k)f^*(k,\phi) \\
    N_u(k)(\omega_{u}(k)+2J_{AA}\cos k) & N_l(k)(\omega_{l}(k)+2J_{AA}\cos k)
\end{array}
\right)
\end{align}
where $N_{u/l}(k)$ is a normalization factor
\begin{equation}
    N_{u/l}(k) = \frac{1}{\sqrt{|f(k,\phi)|^2+(\omega_{u/l}(k)+2J_{AA}\cos k)^2}}.
\end{equation}

\section{Single-qubit self-energy}\label{app:single_qubit_selfenergy}

To compute $c_e^D(t)$, we use the resolvent operator method~\cite{Cohen-Tannoudji1992}, which tell us that the probability amplitude can be computed as:
\begin{equation}\label{eq:c_resolvent}
c_e^D(t)=-\frac{1}{2\pi i}\int_{-\infty}^{\infty} dE\, G_e^D(E+i0^+)e^{-iEt},
\end{equation}
where $G_e^D(z)$ is the single-qubit Green function when it is coupled to the sublattice $D$:
\begin{equation}
G_e^D(z)=\frac{1}{z-\Delta-\Sigma_e^D(z)},
\end{equation}
being $\Sigma_e^D(z)$ the so-called self-energy. In this Section, we derive the expressions for the single-qubit self-energy when the qubit is locally coupled to $A$ or $B$. The $\Sigma_e^D(z)$ of our two band model reads:
\begin{equation}
    \Sigma_e^D(z)=\sum_k \sum_{\alpha=u,l}\frac{|\langle 0 |\alpha_k H_\text{int}\sigma^+|0\rangle|^2}{z-\omega_\alpha(k)}.
\end{equation}

Considering $H_\text{int}$ (Eq. \eqref{eq:Hint} for a single qubit) coupled to $A$ or $B$ and taking into account the relation between $(u_k,l_k)$ and $(a_k,b_k)$ (see Eqs.~\eqref{eq:modes} and \eqref{eq:Pk}) and the expressions for $\omega_{u/l}(k)$ (see Eq. \eqref{eq:bands}):
\begin{align}
    &\Sigma_e^A(z) = \frac{g^2}{2\pi}\int_{-\pi}^{\pi} dk \frac{z}{z^2+2zJ_{AA}\cos k -|f(k,\phi)|^2},\label{eq:SigmaA} \\
    &\Sigma_e^B(z) =  \frac{g^2}{2\pi}\int_{-\pi}^{\pi} dk \frac{z+2J_{AA} \cos k}{z^2+2zJ_{AA}\cos k -|f(k,\phi)|^2}.\label{eq:SigmaB}
\end{align}
We take here the thermodynamic limit: $N\to\infty$. One can solve these integrals by means of the change of variable $y\equiv e^{ik}$. The integration domain is now the unit circle in the complex plane:
\begin{align}
\label{eq:SigmaAint}   &\Sigma_e^A(z) = \frac{g^2}{2\pi i} \oint dy \frac{zJ_{AA}}{(zJ_{AA}-J_{AB}^2e^{i\phi})(y-y_+)(y-y_-)}, \\
\label{eq:SigmaBint}   &\Sigma_e^B(z) = \frac{g^2}{2\pi i} \oint dy \frac{J_{AA}y^2 + zy + J_{AA}}{(zJ_{AA}-J_{AB}^2e^{i\phi})\, y(y-y_+)(y-y_-)},
\end{align}
where $y_\pm$ are:
\begin{equation}\label{eq:ypm}
    y_\pm = \frac{2J_{AB}^2-z^2\pm\sqrt{(2J_{AB}^2-z^2)^2-4(z^2J_{AA}^2+J_{AB}^4-2zJ_{AA}J_{AB}^2\cos\phi)}}{2(zJ_{AA} -J_{AB}^2e^{i\phi})}.
\end{equation}
We define $y_\text{min/max}$ as the minimum/maximum of $\{y_-,y_+\}$ with respect to the absolute values $|y_\pm|$. Applying the Cauchy's residue theorem and taking into account that $(|y_+|-1)(|y_-|-1)<1$ for all $z\in \mathbb{C}$ with $\text{Im}(z)\neq 0$:
\begin{align}
    &\Sigma_e^A(z)=\frac{g^2z\,\text{sign}(|y_-|-|y_+|)}{(zJ_{AA}-J_{AB}^2 e^{i\phi})(y_+-y_-)},\label{eq:SigmaAsol} \\
    &\Sigma_e^B(z) = \frac{g^2J_{AA}}{zJ_{AA}-J_{AB}^2 e^{i\phi}}\left(\frac{1}{y_+y_-} + \frac{y_{\text{min}}^2 + (z/J_{AA}) y_{\text{min}}+1}{y_{\text{min}}(y_{\text{min}}-y_{\text{max}})} \right).\label{eq:SigmaBsol}
\end{align}
\emph{E.g.} if we consider that $\Delta$ is embedded in the lower band, it is straightforward to derive Eq.~\eqref{eq:gammaB} from Eqs.~\eqref{eq:SigmaAsol} and~\eqref{eq:SigmaBsol}.

\section{Qubit-band couplings}\label{app:qubit_band}

Here, we write down the qubit-band coupling for both bands.

Let us consider the interaction Hamiltonian $H_\text{int}$ (Eq. \eqref{eq:Hint}) for a single qubit. For the sake of simplicity, the qubit will be coupled to $A$. We write $H_\text{int}^A$ in terms of $u_k$ and $l_k$ (see Eq. \eqref{eq:modes}):
\begin{align}
& H_\text{int}^A = \frac{g}{\sqrt{N}} \sigma^+ \sum_k e^{ikx_0} (\cos(\theta_k)e^{i\varphi_k}\hat{u}_k + \sin(\theta_k)e^{i\phi_k}\hat{l}_k)+ \text{H.c.} \label{HintA}
\end{align}
where $x_0$ is the position of the qubit and $\cos/\sin(\theta_k)$ are the matrix elements of the unitary transformation $P_k$ (see Eq. \eqref{eq:Pk}). The latter determines the coupling strength to each band: $G_{u,A}(k) = |\cos(\theta_k)|^2$ and $G_{l,A}(k) = |\sin(\theta_k)|^2$, up to the density of states, which is given by $1/|\partial\omega_{u/l}(k)|$.

\section{Trivial crossing point}

The existence of a nonanalytical point in the band structure can cause exotic behaviour both in the dynamics of the emitters and in the effective interactions \cite{gonzaleztudela18c}. We explain in this section why the apparent kink in the bands at $k=\pm\pi/2$ when $\phi=\pm\pi/2$ (see Fig. \ref{fig:sketch}(c) for $\phi=\pi/2$) is actually a trivial crossing point.

\begin{figure}[tbh!]
	\centering
	\includegraphics[scale=0.28]{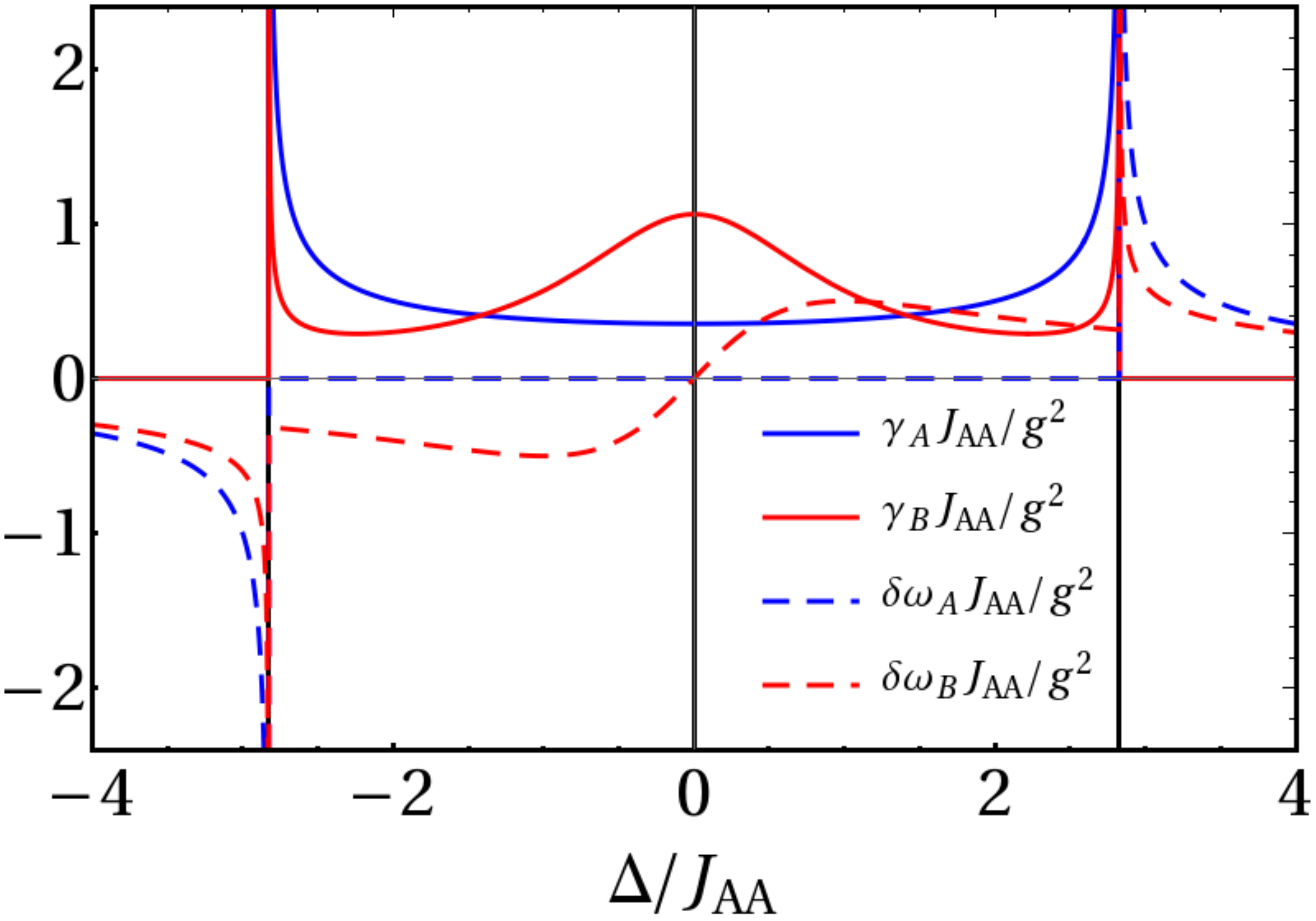}
	\caption{(Color online). Decay rate $\gamma_D = -\text{Im}\Sigma_e^D(\Delta+i0^+)$ (solid lines) and energy shift $\delta\omega_D(\Delta)= \text{Re}\Sigma_e^D(\Delta+i0^+)$ (dashed lines) as a function of the excitation energy of the qubit for the $A$ (blue) and $B$ (red) sublattices. The parameters are $J_{AB}=J_{AA}$ and $\phi=\pi/2$. The vertical black lines stand for the band limits.}
	\label{fig:self_energy_1qb_crossing}
\end{figure}

We first plot both the real and imaginary parts of the self-energy $\Sigma_e^D(z)$, Eqs. \eqref{eq:SigmaAsol} and \eqref{eq:SigmaBsol}, as a function of $\Delta$ when $\phi=\pi/2$ in Fig. \ref{fig:self_energy_1qb_crossing}. As seen, both are smooth functions of $\Delta$, which indicates that the couped qubit does not feel a nonanalyticity in the dispersion relations $\omega_{u/l}(k)$.

\begin{figure}[tbh!]
    \centering
    \includegraphics[scale=0.28]{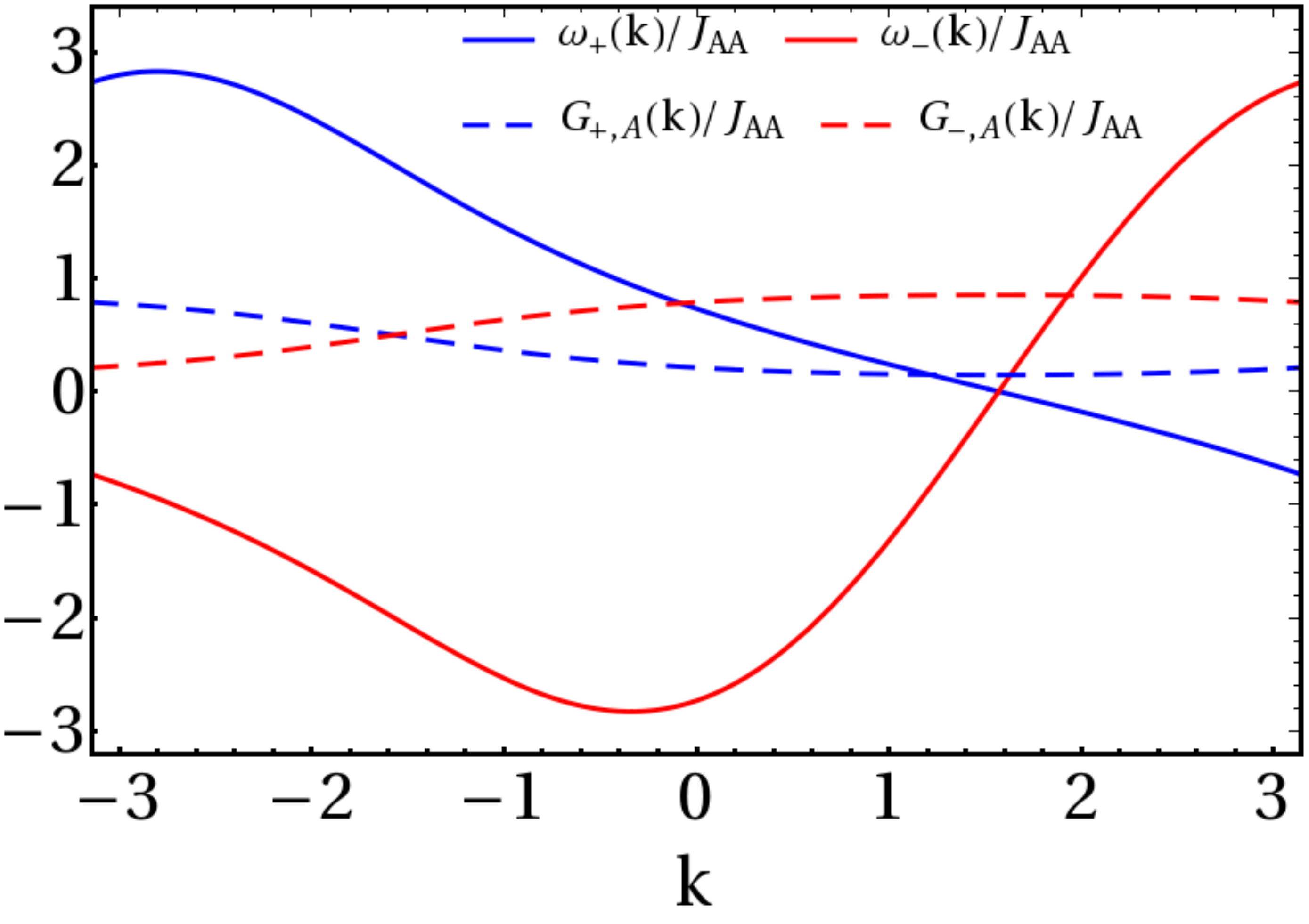}
    \caption{(Color online). Coupling $G_{\pm,A}(k)$ between a qubit placed at sublattice $A$ and $\omega_\pm(k)$ for $\phi=\pi/2$ and $J_{AA}=J_{AB}$. As seen, the couplings $G_{\pm,A}(k)$ are not discontinuous at $k=\pi/2$.}
    \label{fig:coupling_band_pm}
\end{figure}

We confirm this by studying how the QE couples to the bands for $\phi=\pm\pi/2$. We define two new bands $\omega_\pm(k)$, together with the corresponding couplings $G_{\pm,A}(k)$ (see Sect. \ref{app:qubit_band}):
\begin{align}
    \omega_{\pm}(k) \equiv & \left\{
    \begin{array}{c}
        \omega_{u/l}(k) \quad \text{if}\; k<\pi/2,  \\
        \omega_{l/u}(k) \quad \text{if}\; k>\pi/2.
    \end{array}
    \right. \\
    G_{\pm,A}(k) \equiv & \left\{
    \begin{array}{c}
        G_{u/l,A}(k) \quad \text{if}\; k<\pi/2,  \\
        G_{l/u,A}(k) \quad \text{if}\; k>\pi/2.
    \end{array}
    \right.
\end{align}
We plot both $\omega_\pm(k)$ and $G_{\pm,A}(k)$ in Fig. \ref{fig:coupling_band_pm} for $\phi=\pi/2$. As seen, these bands $\omega_\pm(k)$ do not have any kink and the QE couples smoothly to both of them. In conclusion, the apparent nonanalytical behavior is actually an artifact of the definition of the bands.

\section{Sublattice-dependent directional emission \label{SM:sublat}}

\begin{figure}[tbh!]
    \centering
    \includegraphics[scale=0.3]{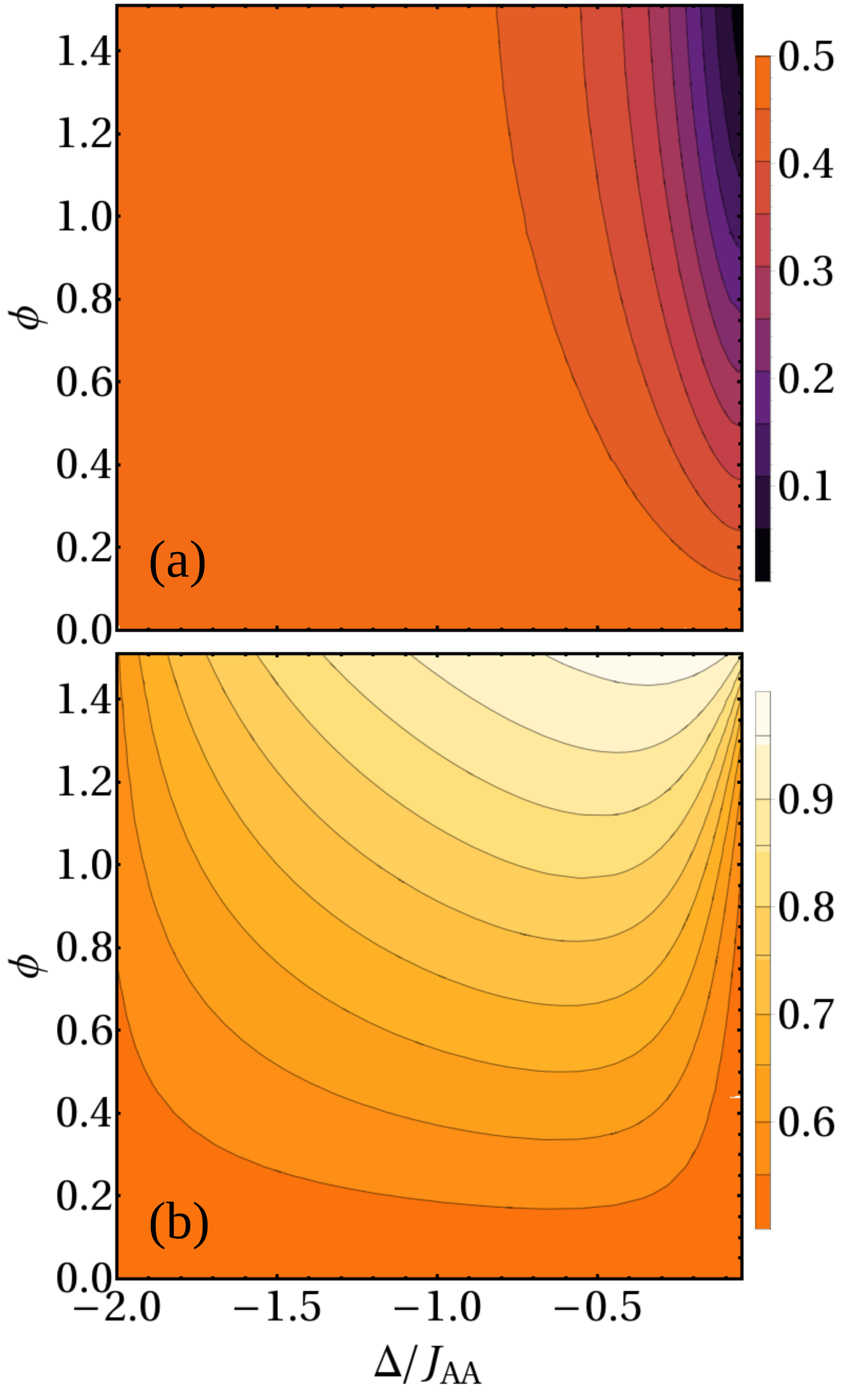}
    \caption{(Color online). Directionality ratio $R_{L,a/b}^{A}$ in panels (a) and (b) respectively for $J_{AB}=0.2 J_{AA}$. Notice that the scale of both panels is different: $(0,0.5)$ in (a) and $(0.5,1)$ in (b).}
    \label{fig:directionality_AB}
\end{figure}

We showed in the main text that a qubit coupled to $B$ emits left-moving photons when $\Delta$ is close to the upper limit of $\omega_l(k)$, $\phi\to\pi/2$, and $J_{AB}\ll J_{AA}$. (see Fig. \ref{fig:directionality}). If it is coupled to $A$ instead, the behaviour changes. We show in Fig. \ref{fig:directionality_AB} the directionality ratios in each sublattice when the qubit is coupled to $A$, $R_{L,a/b}^A$, as a function of $\Delta$ and $\phi$ for $J_{AA}=1$ and $J_{AB}=0.2$. The qubit still decays into left-moving $b$ modes when $\Delta$ is close the upper band limit of $\omega_l(k)$ and $\phi\to\pi/2$, whereas it emits \emph{right}-moving $a$ photons under the same conditions (notice that $R_{L,b}^A$ tends to $1$ in this regime, whilst $R_{L,a}^A\to 0$). This is possible because, in this regime, the matrix elements of $P_k$ are such that the numerator of the couplings $\Gamma_{a,b}^A(k)$  (see Eqs.~\eqref{eq:lambdaa}-\label{eq:lambdab}) compensates the sharp difference between the density of states for $k_L$ and $k_R$. Actually, the global directionality ratio $\eqref{eq:directionality_ratio}$ is totally symmetric: $R_L^A=R_R^A=1/2$, which is straightforwardly derived from the expressions \eqref{eq:lambdaa}, \eqref{eq:lambdab}, \eqref{eq:Pk}, and \eqref{eq:bands}. We do not know whether this can be related to any symmetry of the model and we leave this potential connection for future projects.

\begin{figure}[tbh!]
    \centering
    \includegraphics[scale=0.3]{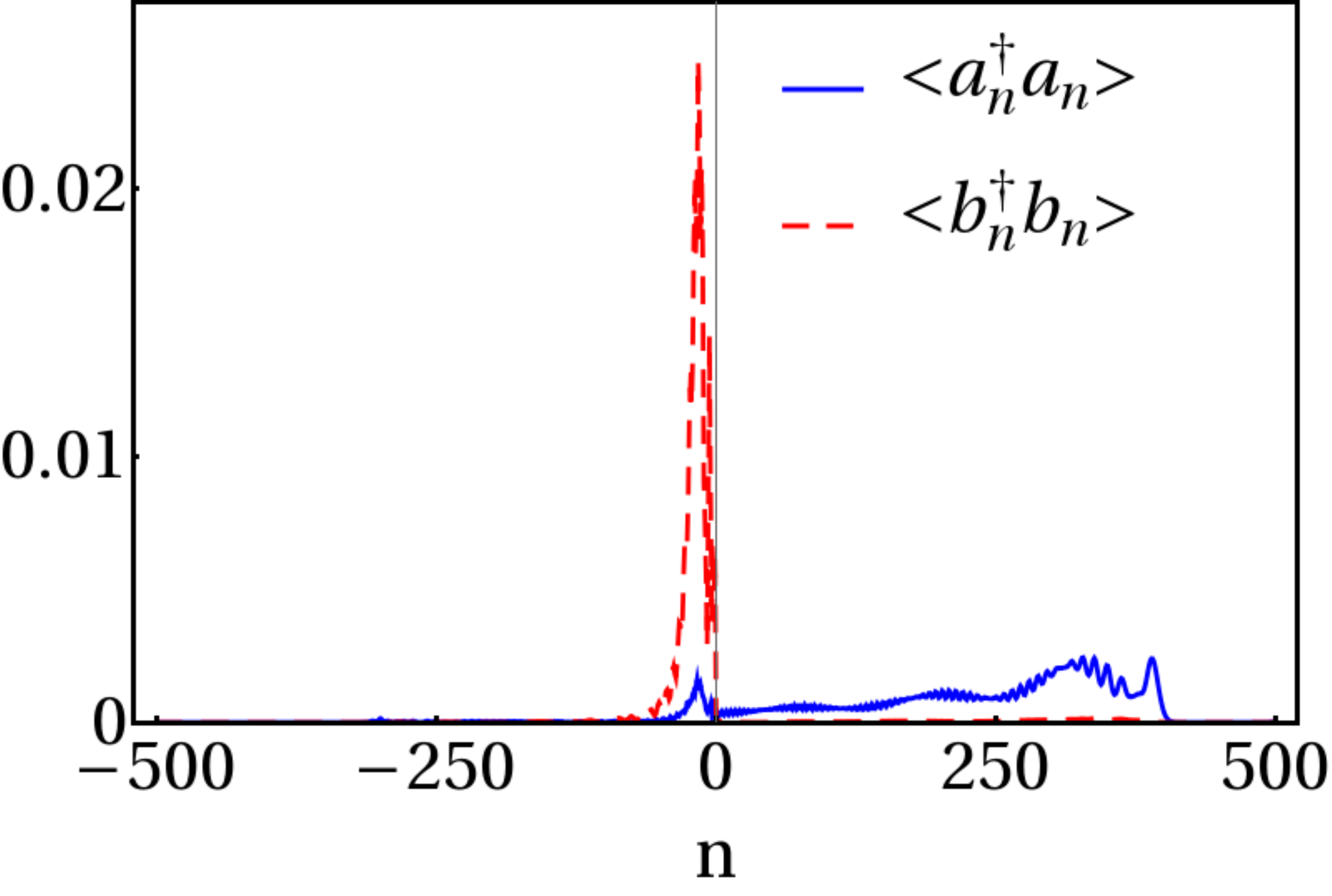}
    \caption{(Color online). Emitted wavepacket when the qubit is coupled to $A$ for $J_{AB}=0.2 J_{AA}$, $\phi=1.5$, $\Delta=-0.1J_{AA}$, $g=0.1 J_{AA}$, $N=1000$, and $t J_{AA}=200$.}
    \label{fig:directionality_AB_wavepackets}
\end{figure}

We show an instance of emission into opposite directions in Fig. \ref{fig:directionality_AB_wavepackets}.

\section{Bound states}\label{app:bs}

In this section, we discuss the existence conditions of the bound states (BSs) and we compute their wavefunctions in real space.

As mentioned in the main text, we have to impose the eigenvalue equation $H\ket{\Psi_m^D} = E_m^D\ket{\Psi_m^D}$ with the energy $E_m^D$ outside of the bands. This can be mapped into finding the roots of the following function: 
\begin{equation}
F_D(E)\equiv E-\Delta- \Sigma_e^D(E),
\end{equation}
with $E \notin \omega_{l,u}(k)$~\cite{Shi2016}. It can be easily proved from Eq.~\eqref{eq:sigmaD} that $\Sigma_e^D(E)$ is a decreasing function, so $F_D(E)$ is an increasing function. Besides, $\lim_{E\to\pm\infty}F_D(E)=\pm\infty$. Then, according to the behaviour of $\Sigma_e^D(E)$ in the band edges, we can figure out whether there exist or not a bound state in each of the band gaps:

\begin{itemize}
\item A BS exists with $E_\text{bs}<\omega_l(k)$ ($E_\text{bs}>\omega_l(k)$) for all the values of the parameters if and only if $F_D(E)>(<)0$ when $E$ tends to the minimum of $\omega_l(k)$ (maximum of $\omega_u(k)$).  We plot $\Sigma_e^D(E)$ for $E$ outside of the band in Fig.~\ref{fig:self_energy_1qb} and show that it diverges in the the lowest/highest energy band-edge, which guarantees the aforementioned conditions, so the existence of two BS below $\omega_l(k)$ and over $\omega_u(k)$, which we label as $\ket{\Psi_{-1}^D}$ and $\ket{\Psi_{+1}^D}$, respectively.

\item The situation is different in the middle band-gap. In Fig.~\ref{fig:self_energy_1qb}(a), we observe when the QE is coupled to the B lattice, the self-energy diverges in both the upper/lower middle band-edges. Thus, a middle BS, $\ket{\Psi_0^B}$, always exists. On the other hand, if the emitter is coupled to $A$, the state $\ket{\Psi_0^A}$ exists if $\Delta > 0$ when $|\phi| < \pi/2$ because the self-energy $\Sigma_e^A(E)$ vanishes when $E$ tends to the maximum of $\omega_l(k)$. When $\phi\in(-\pi,-\pi/2)$ or $\phi\in(\pi/2,\pi)$, the existence condition is $\Delta<0$ (not shown).
\end{itemize}



Concerning the wavefunctions, if $D=A$, the coefficients read
\begin{align}
        c_{m,a}^A(n) = \frac{gc_e^A}{2\pi}\int_{-\pi}^\pi dk\, e^{ikn} & \left(\frac{|(P_k)_{11}|^2}{E_m^D-\omega_u(k)} \right. \nonumber \\
        &+ \left. \frac{|(P_k)_{12}|^2}{E_m^A-\omega_l(k)}\right), \label{eq:caA_P} \\
    c_{m,b}^A(n)  = \frac{gc_e^A}{2\pi}\int_{-\pi}^\pi dk\, e^{ikn} & \left(\frac{(P_k)_{21}(P_k)_{11}^*}{E_m^A-\omega_u(k)} \right. \nonumber\\
        & + \left. \frac{(P_k)_{22}(P_k)_{12}^*}{E_m^A-\omega_l(k)}\right) , \label{eq:cbA_P}
\end{align}
where $E_m^A$ is the energy of $\ket{\Psi_\text{m,bs}^A}$ and $c_e^A$ is obtained imposing the normalization condition. Doing the maths:
\begin{align}
    &c_{m,a}^A(n) = \frac{gc_e}{2\pi}\int_{-\pi}^\pi dk \frac{e^{ikn}\,E_\text{m}^A}{(E_\text{m}^A)^2+J_{AA} 2E_\text{m}^A\cos k -|f(k,\phi)|^2}, \label{eq:caA}
    \\
    &c_{m,b}^A(n) = \frac{-gc_e}{2\pi}\int_{-\pi}^\pi dk \frac{e^{ikn}f^*(k,\phi)}{(E_\text{m}^A)^2+2J_{AA} E_\text{m}^A\cos k -|f(k,\phi)|^2}. \label{eq:cbA}
\end{align}
If the qubit is instead coupled to $B$:
\begin{align}
        c_{m,a}^B(n) = \frac{gc_e^B}{2\pi}\int_{-\pi}^\pi dk\, e^{ikn}&\left(\frac{(P_k)_{11}(P_k)_{21}^*}{E_m^D-\omega_u(k)} \right. \nonumber \\
        &+ \left.\frac{(P_k)_{12}(P_k)_{22}^*}{E_m^B-\omega_l(k)}\right), \label{eq:caB_P} \\
    c_{m,b}^B(n)  = \frac{gc_e^B}{2\pi}\int_{-\pi}^\pi dk\, e^{ikn} & \left(\frac{|(P_k)_{21})|^2}{E_m^B-\omega_u(k)} \right. \nonumber\\
    & \left.+ \frac{|(P_k)_{22})|^2}{E_m^B-\omega_l(k)}\right) , \label{eq:cbB_P}
\end{align}
which becomes:
\begin{align}
    &c_{m,a}^B(n) = \frac{-gc_e^B}{2\pi}\int_{-\pi}^\pi dk \frac{e^{ikn}f(k,\phi)}{(E_\text{m}^B)^2+2J_{AA} E_\text{m}^B\cos k -|f(k,\phi)|^2}, \label{eq:caB}\\
    &c_{m,b}^B(n) = \frac{gc_e^B}{2\pi}\int_{-\pi}^\pi dk \frac{e^{ikn}(E_\text{m}^B+2 \cos k)}{(E_\text{m}^B)^2+2J_{AA} E_\text{m}^B\cos k -|f(k,\phi)|^2}. \label{eq:cbB}
\end{align}
Notice that all these expressions look similar to $\Sigma_e^D(z)$ (see Eqs. \eqref{eq:SigmaA} and \eqref{eq:SigmaB}), so we can calculate the coefficients in terms of complex integrals (Eqs. \eqref{eq:SigmaAint} and \eqref{eq:SigmaBint}). The change of variable is still $y=e^{ik}$ if $n\geq 1$, but $y=e^{-ik}$ if $n\leq -1$. In the first case, the poles of the integral are $y_\pm$, while in the second are their complex conjugates $y_\pm^*$.


\section{Two-qubit self-energy}\label{app:two_qubit_selfenergy}

We show here the expressions for the collective self-energy $\Sigma_{c}^{D_{12}}$ (Eq. \eqref{eq:manybody}). The computation is totally analogous to the single-qubit self-energy (see App. \ref{app:single_qubit_selfenergy}). They read
\begin{align}
    &\Sigma_{c}^{AA}(z;r_{12}) = \frac{g^2}{2\pi}\int_{-\pi}^\pi dk \frac{e^{ikr_{12}}\,z}{z^2+2zJ_{AA}\cos k -|f(k,\phi)|^2}, \label{eq:sigma12aa} \\
    &\Sigma_{c}^{BB}(z;r_{12}) = \frac{g^2}{2\pi}\int_{-\pi}^\pi dk \frac{e^{ikr_{12}}(z+2J_{AA}\cos k)}{z^2+2zJ_{AA}\cos k -|f(k,\phi)|^2}, \label{eq:sigma12bb}\\
    &\Sigma_{c}^{AB}(z;r_{12}) = -\frac{g^2}{2\pi}\int_{-\pi}^\pi dk \frac{e^{ikr_{12}}f^*(k,\phi)}{z^2+2zJ_{AA}\cos k -|f(k,\phi)|^2}, \label{eq:sigma12ab}
\end{align}
where $r_{12}=x_2-x_1$ is the relative position of the qubits. Notice that $\Sigma_{c}^{AA}(z;r_{12})$, $\Sigma_{c}^{BB}(z;r_{12})$, and $\Sigma_{c}^{AB}(z;r_{12})$ are proportional to the bound-state coefficients $c_{m,a}^A(r_{12})$, $c_{m,b}^B(r_{12})$, and $c_{m,b}^A(r_{12})$ respectively, by changing the bound-state energies $E_m^D$ by $z$ (see Eqs. \eqref{eq:caA}, \eqref{eq:cbB}, and \eqref{eq:cbA}). It is here where it becomes evident that the effective interactions are mediated by the bound states.

Finally, we can compute the accumulated phase of a closed loop in the effective spin lattice. \emph{E.g.} taking the parameters of Fig. \ref{fig:bound_state_mid} ($J_{AA}=J_{AB}=1$, $\phi=2.094$, $\Delta=-0.01$, and $g=0.1$) and choosing the closed path $a\to a\to b\to a$, this phase is $\text{arg}(\Sigma_c^{AA}(\Delta;1))+\text{arg}(\Sigma_c^{AB}(\Delta;1))+\text{arg}(\Sigma_c^{AB}(\Delta;-1))\simeq -1.22$. As it is nonzero, the effective models can simulate systems without time and parity invariance.

\bibliographystyle{apsrev4-1}
\bibliography{bib_ST,Sci}

\end{document}